\documentclass[12pt,preprint]{aastex}
\usepackage{epsfig}

\def\degree{$^\circ$}
\def\arcs#1{$#1''$}
\def\arcsa#1#2{$#1^{\prime\prime}_{^\textrm{.}}#2$}

\def\solarmass{$M_\odot$}
\def\solarmasse{M_\odot}

\def\uJyb{$\mu$Jy beam$^{-1}$}

\def\cmc{cm$^{-3}$}
\def\cms{cm$^{-2}$}
\def\micron{$\mu$m}

\def\mHt{m_{\textrm{\scriptsize H}_2}}

\def\Ro{R_\textrm{\scriptsize 0}}

\def\To{T_\textrm{\scriptsize 0}}
\def\no{n_\textrm{\scriptsize 0}}

\def\H2{H$_2$}
\def\N2HP{N$_2$H$^+$}

\def\NH3{NH$_3$}

\def\mH2{m_{\textrm{\scriptsize H}_2}}
\def\no{n_\mathrm{o}}
\def\na{n_\mathrm{t}}
\def\Ro{R_\mathrm{o}}
\def\Ra{R_\mathrm{t}}
\def\To{T_\mathrm{o}}
\def\Ta{T_\mathrm{t}}
\def\ho{h_\mathrm{o}}
\def\ha{h_\mathrm{t}}

\def\vk{v_\mathrm{ko}}
\def\cs{c_\mathrm{s}}
\def\vko{v_\mathrm{ko}}
\def\cso{c_\mathrm{so}}
\def\vp{v_\phi}

\def\iI{{\it I}}
\def\iQ{{\it Q}}
\def\iU{{\it U}}
                                                                            
\def\putfiga#1#2#3{\epsfig{scale=#1,angle=#2,figure=#3}}
\def\putfig#1#2#3{}
\def\leftblank#1{}

\begin{document}

\title{Polarization Substructure  in the Spiral-Dominated HH 111 Disk: Evidence for Grain Growth}


\author{Chin-Fei Lee\altaffilmark{1}, Zhi-Yun Li\altaffilmark{2}, Tao-Chung
Ching\altaffilmark{3}, Haifeng Yang\altaffilmark{4}, Shih-Ping
Lai\altaffilmark{5}, Zhe-Yu Daniel Lin\altaffilmark{2}, Ying-Chi
Hu\altaffilmark{1,5}}

\altaffiltext{1}{Academia Sinica Institute of Astronomy and Astrophysics,
No.  1, Sec.  4, Roosevelt Road, Taipei 106216, Taiwan, R.O.C.;
cflee@asiaa.sinica.edu.tw}
\altaffiltext{2}{Astronomy Department, University of Virginia, Charlottesville, VA 22904}
\altaffiltext{3}{National Radio Astronomy Observatory, P.O. Box O, Socorro, NM 87801, USA}
\altaffiltext{4}{Institute for Astronomy, School of Physics, Zhejiang University, Hangzhou, 310027 Zhejiang, China}
\altaffiltext{5}{Institute of Astronomy and Department of Physics, National
Tsing Hua University, Hsinchu, Taiwan}

\begin{abstract}

The HH 111 protostellar disk has recently been found to host a pair of
spiral arms.  Here we report the dust polarization results in the disk as
well as the inner envelope around it, obtained with the Atacama Large
Millimeter/submillimeter Array in continuum at $\lambda \sim$ 870 \micron{}
and $\sim$ \arcsa{0}{05} resolution.  In the inner envelope, polarization is
detected with a polarization degree of $\sim$ 6\% and an orientation almost
everywhere parallel to the minor axis of the disk, and thus likely to be due to
the dust grains magnetically aligned mainly by toroidal fields.  In the disk, the
polarization orientation is roughly azimuthal on the far side and becomes
parallel to the minor axis on the near side, with a polarization gap in
between on the far side near the central protostar.  The disk polarization
degree is $\sim$ 2\%.  The polarized intensity is higher on the near side
than the far side, showing a near-far side asymmetry.  More importantly, the
polarized intensity and thus polarization degree are lower in the spiral
arms, but higher in between the arms, showing an anticorrelation of the
polarized intensity with the spiral arms.  Our modeling results indicate
that this anticorrelation is useful for constraining the polarization
mechanism and is consistent with the dust self-scattering by the grains that
have grown to a size of $\sim$ 150 \micron{}.  The interarms are sandwiched
and illuminated by two brighter spiral arms and thus have higher polarized
intensity.  Our dust self-scattering model can also reproduce the observed
polarization orientation parallel to the minor axis on the near side and the
observed azimuthal polarization orientation at the two disk edges in the
major axis.  Further modeling work is needed to study how to reproduce the
observed near-far side asymmetry in the polarized intensity and the observed
azimuthal polarization orientation on the far side.


\end{abstract}


\section{Introduction}

With the powerful Atacama Large Millimeter/submillimeter Array (ALMA), dust
polarization has been detected towards quite a few protostellar disks in
millimeter and submillimeter wavelengths over the past decade
\cite[e.g.,][]{Kataoka2017,Stephens2017,Stephens2023,Lee2018Pdisk,Lee2021Pdisk,
Bacciotti2018,Cox2018,Girart2018,Harris2018,Hull2018,Dent2019,Sadavoy2019,Tang2023,Lin2024,Yang2024}. 
It could be due to magnetically aligned grains \citep{Andersson2015}, dust
self-scattering \citep{Kataoka2015,Yang2016}, or radiatively aligned grains
\citep{Kataoka2017,Tazaki2017}, and more than one mechanism can be at work
simultaneously, perplexing the analysis.  If produced by magnetically
aligned grains, it can be used to infer magnetic field morphology in the
disks, which is crucial to study the launching of the jets and winds from
the disks \citep{Konigl2000,Shu2000} and disk accretion \citep{Turner2014}. 
If produced by dust self-scattering, it can be used to infer the grain size
in the disks, allowing us to study whether there is a grain growth leading
ultimately to planet formation.

In the protostellar disks, substructures, such as rings, gaps, and spirals,
have also been detected, and may help narrow down the dust polarization
mechanism because they can affect dust polarization differently for
different polarization mechanisms.  In particular, the dust polarization due
to dust self-scattering is affected by the asymmetry in the flux incoming
from various directions, while the dust polarization due to the magnetically
aligned grains is not.  Recent dust polarization observations towards the
well-studied HL Tau disk at submillimeter wavelength of 870 \micron{} have
indeed shown that the rings and gaps in the disk have different
polarization degree and orientation, which can be attributed to both
scattering and emission from the aligned grains \citep{Stephens2023,Lin2024}.

The HH 111 disk is a young ($\sim 5\times10^5$ yr) Class I
protostellar disk located in Orion at a distance of $\sim$ 400 pc and has
recently been found to harbor a pair of trailing spiral arms
\citep{Lee2020HH111}.  Previous molecular line observations have found it to
have a radius of $\sim$ \arcsa{0}{4} (160 au) and a Keplerian rotation
\citep{Lee2016HH111}.  It drives the powerful collimated jet HH 111
\citep{Reipurth1999}, suggestive of a need for a poloidal magnetic field at
the center of the disk according to the current popular jet launching models
\citep{Shu2000,Konigl2000}.  The disk is close to edge-on, optimal for
searching for poloidal field in the disk.  Dust polarization has been
detected toward the disk at $\sim$ \arcsa{0}{12} resolution
\citep{Lee2018Pdisk} at $\lambda \sim$ 870 \micron{}, but showing a complex
polarization pattern that could be due to either magnetically aligned grains
or dust self-scattering.  In order to better determine the origin of the
dust polarization, here we report the dust polarization detection at 2.4
times as high resolution and the modeling with the two polarization
mechanisms.  At this higher resolution, we can clarify the polarization
pattern and its morphological relationship with the spiral arms.  In
addition, the spiral arms have been argued to be excited by gravitational
instability to facilitate the disk accretion \citep{Lee2020HH111}, allowing
us to constrain the disk density and thus the dust absorption opacity in the
disk.  By modeling the better resolved dust polarization, we can better
determine the polarization mechanisms, searching for poloidal field for jet
launching and grain growth for planet formation.

\section{Observations}\label{sec:obs}

Polarization observations toward the HH 111 system were carried out with
ALMA in Band 7 in Cycle 3 (Project ID: 2015.1.00037.S) in 2016 and Cycle 5
(Project ID: 2017.1.00044S) in 2017.  Since the details of the observations
in Cycles 3 and 5 have been reported in \citet{Lee2018Pdisk} and
\citet{Lee2020HH111}, respectively, we only recap the observation
information most directly relevant to our investigation.  In these
observations, one pointing was used toward the center of the HH 111 system,
with a field of view (the primary beam) having a size of $\sim$
\arcsa{17}{8}.  The correlator was set up to have four continuum windows,
centered at 336.5, 338.5, 348.5, and 350.5 GHz for a total bandwidth of
$\sim$ 8 GHz centered at 343.5 GHz (corresponding to $\lambda \sim$ 870
\micron).  The projected baseline lengths were $\sim$ 14-8550 m, after
combining the observations in the two cycles.  The maximum recoverable size
(MRS) scale was $\sim$ \arcsa{1}{4}, enough to cover the disk seen in the
continuum.



In both cycles, the $uv$ data were calibrated with the Common Astronomy
Software Applications (CASA) package, with quasars J5010+1800, J0552+0313,
and J0522-3627 as bandpass, gain, and polarization calibrators,
respectively.  We also performed a phase-only self-calibration to improve
the map fidelity.  The calibrated data from the two cycles were then
combined into one calibrated data for mapping.  In order to map the disk
structure and dust polarization with increasing details, we used 3
decreasing robust factors for the $uv$ weighting in making the maps.  The
robust factors are 2, 0.5, and 0.  The resulting resolutions are
\arcsa{0}{080}$\times$\arcsa{0}{070}, \arcsa{0}{062}$\times$\arcsa{0}{046},
and \arcsa{0}{041}$\times$\arcsa{0}{030}, respectively.  In the resulting
Stokes \iI{} maps, the noise levels $\sigma$ are $\sim$ 100 \uJyb{} (186
mK), 80 \uJyb{} (290 mK), and 50 \uJyb{} (417 mK), respectively.  In the
resulting Stokes \iQ{} and \iU{} maps, the noise levels $\sigma_p$ are
$\sim$ 16 \uJyb{} (30 mK), 18 \uJyb{} (65 mK), and 26 \uJyb{} (217 mK),
respectively.  The linear polarization intensity is defined as $P_i =
\sqrt{Q^2+U^2-\sigma_p^2}$ and thus is bias-corrected.  The fraction of the
polarization is then defined as $P=P_i/I$.  According to ALMA Technical Handbook,
the instrumental error on $P$ is expected to be $\lesssim$ 0.2\% for the
disk at the phase center, because the disk size is $\lesssim$ \arcs{1} and
thus much smaller than 1/3 of the primary beam.  Polarization orientations
are defined by the $E$ vectors.



\section{Results}

\begin{figure} [!hbp]
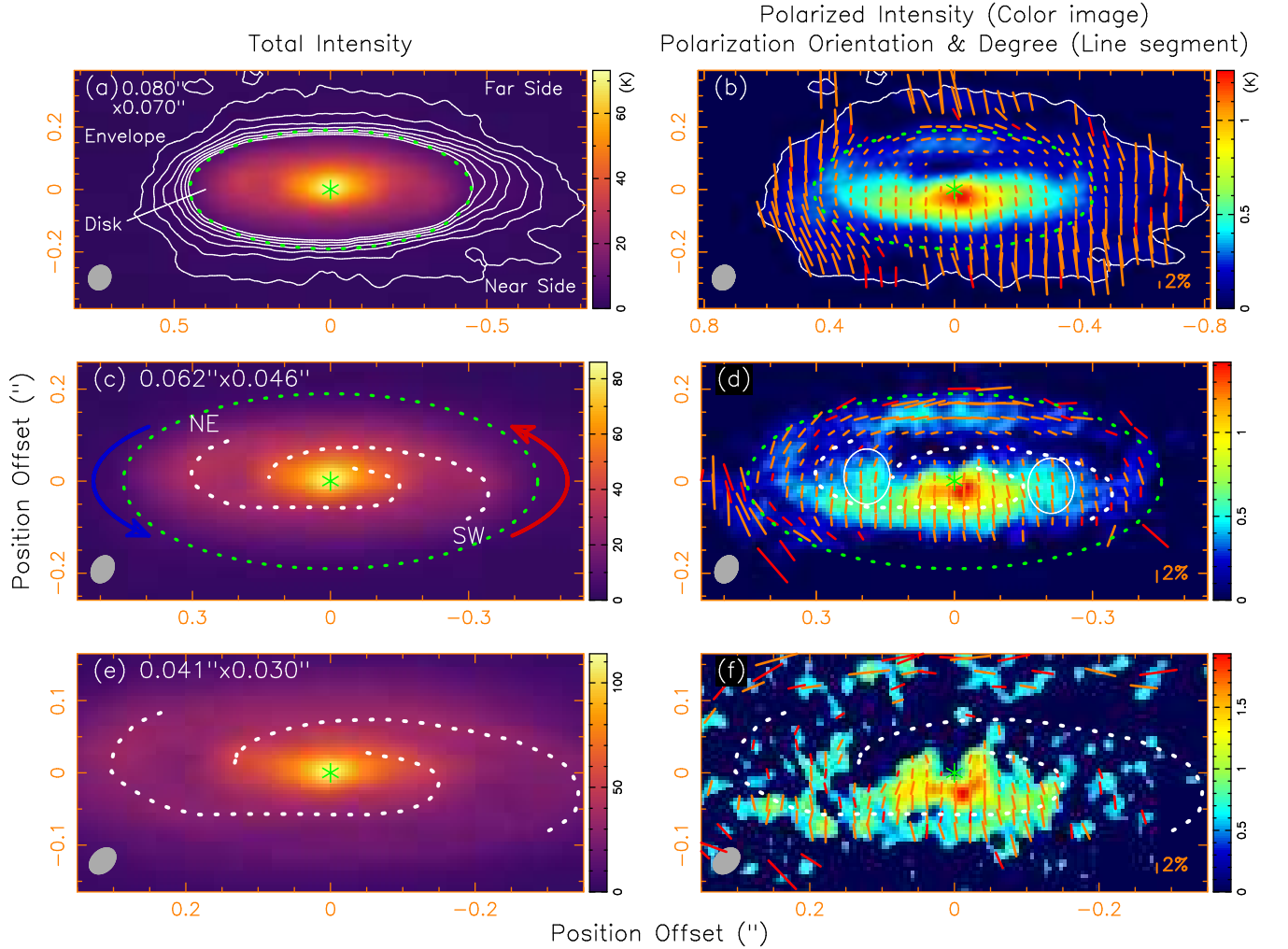

\centering
\putfiga{0.7}{270}{f1.eps} 
\figcaption[]
{Polarization maps of the continuum emission at $\lambda \sim$ 870 \micron{}
toward the HH 111 disk and envelope at angular resolutions of
\arcsa{0}{080}$\times$\arcsa{0}{070}, \arcsa{0}{062}$\times$\arcsa{0}{046},
and \arcsa{0}{041}$\times$\arcsa{0}{030}, respectively, in top, middle, and
bottom row.  Left column shows the total intensity.  Right column shows the
polarized intensity with color image, and polarization orientation and
degree with line segments (red for detections of 2.5$-$3 $\sigma_{p}$ and orange 
for detections greater than 3 $\sigma_{p}$).
(a) Contour levels start from 6 $\sigma$ with a
step of 6 $\sigma$, where $\sigma$=186 mK.  (b) The contour shows the first
contour shown in (a).  In (a)-(d), the green dotted ellipse marks the
rough boundary of the disk.  In (c)-(f), the white dashed curves mark the
spiral arms previously identified in \citet{Lee2020HH111}.  In (d), the two
white ellipses show the interarm regions with high polarized intensity.
\label{fig:obsmap}}
\end{figure}

Figure \ref{fig:obsmap} shows the maps of total intensity, polarized
intensity, polarization orientation and degree of the continuum emission
observed toward the center of the HH 111 protostellar system at $\lambda
\sim$ 870 \micron{}.  At this submillimeter wavelength, the continuum traces
the dust emission in the disk and inner envelope \citep{Lee2009HH111}.  The maps
have been rotated by 83\degree{} counterclockwise to align the disk major
axis with the horizontal axis.  They are presented from low ($\sim$
\arcsa{0}{075} in Figs.  \ref{fig:obsmap}a, b) to medium ($\sim$
\arcsa{0}{054} in Figs.  \ref{fig:obsmap}c, d) and then to high ($\sim$
\arcsa{0}{035} in Figs.  \ref{fig:obsmap}e, f) angular resolution, in order
to reveal the disk structure and dust polarization with increasing details.


At low angular resolution, the continuum emission is detected in both the
disk and the inner part of the envelope around the disk (see Figure
\ref{fig:obsmap}a).  The contours of the continuum emission are crowded at a
radius of $\sim$ \arcsa{0}{45} on the major axis, indicating that the
continuum emission intensity and thus the dust surface density increases rapidly
there, confirming the formation of a disk within the envelope.  The green
dotted ellipse is thus added to mark the rough boundary between the disk and
the envelope.  It is highly elongated, consistent with the disk tilted only
by $\sim$ 18\degree{} away from being edge-on \citep{Lee2020HH111}.  As
delineated by the contour at the lowest level (6$\sigma$ detection), the
inner envelope is not only detected around the two edges of the disk along the major
axis, but also above and below the disk extending slightly away from the disk. 
Interestingly, the polarization map (Figure \ref{fig:obsmap}b) shows a
change of polarization morphology and a decrease of polarization degree from
the envelope to the disk.  In the envelope, the polarization orientation is
almost everywhere parallel to the minor axis, while in the disk, the
polarization orientation is azimuthal on the far side and then becomes
parallel to the minor axis on the near side, with a polarization gap seen in
between at $\sim$ \arcsa{0}{07} above the central protostar.  The
polarization degree is $\sim$ 6\% in the envelope but decreases to $\sim$
2\% in the disk.  Inside the disk, the polarized intensity is higher on the
near side than the far side, producing a near-far side asymmetric
distribution in the polarized intensity.  Near the central source, polarized
emission is mainly detected on the near side.

Figure \ref{fig:obsmap}c zooms in to the disk at medium angular resolution. 
The dotted curves delineate the northeastern (NE) and southwestern (SW)
spiral arms previously identified at a slightly higher angular resolution of
$\sim$ \arcsa{0}{043} in \citet{Lee2020HH111}.  The spiral arms are
trailing, with their outer tip pointing in the direction opposite to the
disk rotation measured before (as indicated by the blue and red curved
arrows).  At this resolution, it becomes clear that the polarized intensity
is high at the two ends of the highly inclined disk, as well as the lower
and upper rims of the disk (see Figure \ref{fig:obsmap}d).  More
importantly, the polarized intensity is lower and even below the detection
in the spiral arms (as marked by the dotted curves), but higher in between
the arms (as marked by the two white ellipses), except for the inner part of
the NE spiral arm on the near side where it is not resolved and thus not
clear.  This suggests an anticorrelation between the spiral arms and the
polarized intensity (and thus the polarization degree).  In order to check
this behavior in the inner part of the NE spiral arm, we also present the
total intensity and polarized intensity maps at high resolution in Figures
\ref{fig:obsmap}e and \ref{fig:obsmap}f, respectively.  As can been seen,
the polarized intensity there in the spiral arm could also be lower than its
surrounding, and observations at higher resolution and sensitivity are
needed to confirm it.

\section{Origin of the Dust Polarization in the Inner Envelope}

\begin{figure} [!hbp]
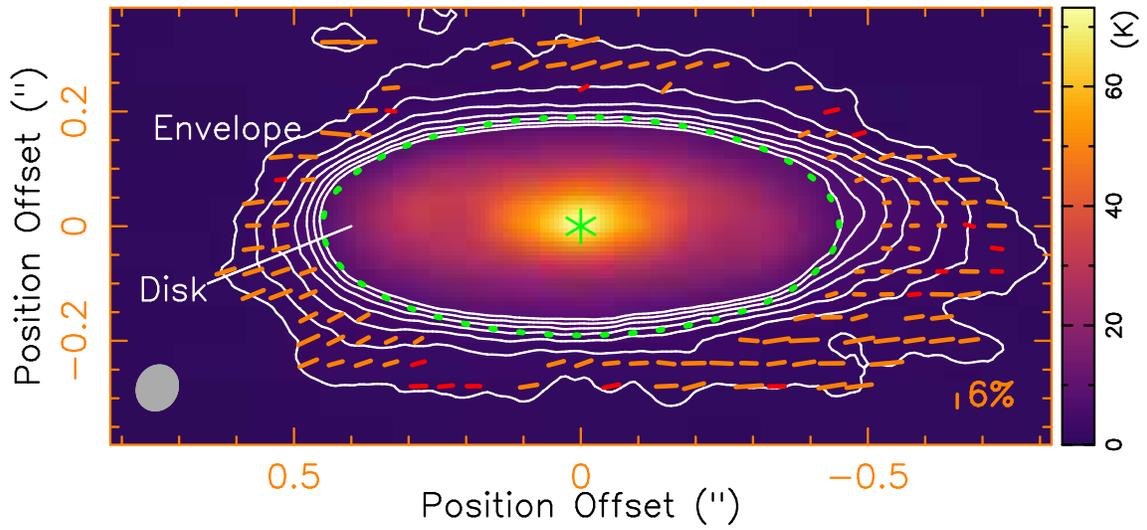

\centering
\putfiga{1.2}{270}{f2.eps} 
\figcaption[]
{Magnetic field orientations in the inner envelope plotted on top of the
total intensity of the continuum map. The color image, contour levels, and green dotted
ellipse are the same as in Figure \ref{fig:obsmap}a. The field orientations
are obtained by rotating the polarization orientations by 90\degree{}.
\label{fig:obsbfield}}
\end{figure}

Dust polarization in the inner envelopes of young protostellar systems
observed at $\lambda \sim$ 870 \micron{} on a few hundred au scale has been
found to be mainly due to magnetically aligned grains \cite[see,
e.g.,][]{Cox2018,Maury2018,Lee2019HH211,Kwon2019}.  For the case of HH 111,
the dust polarization is also likely due to magnetically aligned grains. 
The inner envelope has a brightness temperature of $\sim$ 1$-$6 K, much
lower than the temperature there previously derived from the C$^{18}$O gas,
which is $\sim$ 40 K \citep{Lee2010}.  Therefore, the continuum emission is
optically thin at the observed wavelength.  As a result, the dust
polarization is likely due to thermal emission of the dust and magnetic
field orientation can be obtained by rotating the polarization orientation
by 90\degree{}, as shown in Figure \ref{fig:obsbfield}.  In this case,
the magnetic fields in the inner envelope are almost everywhere parallel to
the major axis and are, thus, mainly toroidal, although a highly pinched
poloidal (particularly radial) component is needed to explain the observed
polarization orientations near the two edges along the major axis where the
toroidal field component is close to the line of sight in this highly
inclined system (and thus contributes little to the polarization). This is
reasonable, because the motion there is dominated by rotation
\citep{Lee2016HH111}, which twists the field lines into a predominantly
toroidal configuration \cite[see, e.g.,][]{Machida2008,Li2013,Kwon2019}.

\section{Origin of the Dust Polarization in the Disk}

Previous study at a lower resolution of $\sim$ \arcsa{0}{12} suggested that
the dust polarization in the HH 111 disk could be due to magnetically
aligned grains, dust self-scattering, or both \citep{Lee2018Pdisk}.  In
order to further determine the origin of the dust polarization in the disk,
we construct a fiducial disk model and then generate dust polarization maps
with these two polarization mechanisms to compare with the dust polarization
mapped here at higher resolutions.

\subsection{Disk Model}


\def\mHt{m_{\textrm{\scriptsize H}_2}}
\def\mH2{m_{\textrm{\scriptsize H}_2}}
\def\no{n_\mathrm{o}}
\def\na{n_\mathrm{t}}
\def\rhoa{\rho_\mathrm{t}}
\def\Ro{R_\mathrm{o}}
\def\zo{z_\mathrm{o}}
\def\Ra{R_\mathrm{t}}
\def\hd{h_\mathrm{d}}
\def\To{T_\mathrm{o}}
\def\Ta{T_\mathrm{t}}
\def\ho{h_\mathrm{o}}
\def\ha{h_\mathrm{t}}
\def\hs{h_\mathrm{s}} 
\def\xo{x_\textrm{\scriptsize 0}}

\def\vk{v_\mathrm{ko}}
\def\cs{c_\mathrm{s}}
\def\vko{v_\mathrm{ko}}
\def\cso{c_\mathrm{so}}
\def\vp{v_\phi}

\def\kabs{\kappa_\textrm{\scriptsize abs}}
\def\ksca{\kappa_\textrm{\scriptsize sca}}

To construct the fiducial model, we adopt the same flared disk model previously used to model the continuum
map of this disk obtained at a similar resolution in \citet{Lee2020HH111}. 
Therefore, the dust in the disk has the following mass
density and temperature distributions in cylindrical coordinates
$(R,\phi,z)$
\begin{eqnarray} 
\rho (R,z) &=& \rhoa
(\frac{R}{\Ra})^{-n} \exp(-\frac{z^2}{2 \hd^2}) \nonumber \\ 
T (R,z) &=& \Ta (\frac{R}{\Ra})^{-q}
\end{eqnarray}
where $\Ra$ is the turnover radius to be defined below, $\rhoa$ and $\Ta$
are the mass density and temperature in the disk midplane at $\Ra$,
respectively, $n$ and $q$ are the power-law index for the density and
temperature, respectively, and $\hd$ is the pressure scale height.    The scale height depends
on the ratio of the sound speed to the angular rotation speed and thus
increases with radius as $R^{(3-q)/2}$.  It can decrease 
near the disk edge and is thus assumed to be given by
\begin{eqnarray}
\hd (R)= \ha
\left\{
\begin{array}{ll} (\frac{R}{\Ra})^{(3-q)/2} & \;\;\textrm{if}\;\; R \leq
\Ra, \\ \sqrt{1-\frac{3}{4}(\frac{R-\Ra}{\Ro-\Ra})^2} & \;\;\textrm{if}\;\;
\Ra < R \leq \Ro 
\end{array} \right.
\end{eqnarray}
so that it increases to $\ha$ at $\Ra$, and then drops to $\ha/2$ at the outer radius $\Ro$.
Moreover, the disk is assumed to have a surface with a height $\ho =
\sqrt{2} \hd$, where the density drops to $1/e$ of that in the midplane.

A pair of trailing spiral arms are also added into the disk.  They are
assumed to follow the logarithmic structure as excited by gravitational
instability (GI) due to the self gravity of the disk \citep{Lee2020HH111},
with the following amplitude
\begin{equation}
s = \cos[m(\phi-\frac{1}{a}\ln (R/R_s)-\phi_0)].
\end{equation}
Here $m=2$ is adopted to produce a pair of spiral arms, $R_s$ is the outer radius of
the spiral arms at $\phi=\phi_0$, and $\tan^{-1} a$ is the pitch angle.  As
seen in the HH 211 disk \citep{Lee2023HH211}, the spiral arms are assumed to
grow from the midplane.  Therefore, the total dust density and temperature
of the disk are assumed to decrease in the vertical direction from the
midplane with the following vertical profile:
\begin{eqnarray}
\rho^s(R,\phi,z) &=& \rhoa (\frac{R}{\Ra})^{-n} \exp(-\frac{z^2}{2 \hd^2})[1+A(1+s) \exp(-\frac{z^2}{2 \hs^2})] \nonumber \\
   T^s(R,\phi,z) &=& \Ta (\frac{R}{\Ra})^{-q}[1+ B(1+s) \exp(-\frac{z^2}{2 \hs^2})]
\end{eqnarray}
where $h_s$ is the scale height of the spiral arms and is assumed to be a
fraction of the disk scale height, i.e., $\hs = f \hd$ with $f < 1$.  $A$
and $B$ are the spiral amplitude in density and temperature, respectively. 
The spiral arms are seen extending from a radius of $\sim$ \arcsa{0}{4} down
to $\sim$ \arcsa{0}{1} from the central protostar \citep{Lee2020HH111}.  It
is unclear how much closer they can extend because the disk structure there
has not been resolved.  Here they are assumed to extend further in with a
small inner radius of \arcsa{0}{03}, which is about a half of the beam size.


\def\xo{x_\textrm{\scriptsize 0}}
\def\Erf{\textrm{Erf}}

Additional physical quantities can be derived to constrain the model and check for consistency.
Assuming that the disk also contains gas with a gas to dust mass ratio of $\sim$ 100,
then the mean surface density (including gas and dust) of the disk at given $R$
can be derived by integrating the density over $z$ and averaging it over
$\phi$:
\begin{eqnarray}
\Sigma (R) & \sim & 100 \int_{-\ho}^{\ho} \int_0^{2\pi} \rho^s (R,\phi,z) \frac{d\phi}{2\pi} \,dz \nonumber \\
&= & 100 \,\rhoa\, (\frac{R}{\Ra})^{-n} \, \sqrt{2 \pi} \hd \Big[\Erf(1)+ A\, K(f) \Big]\nonumber \\
& =& \Sigma_t
\left\{
\begin{array}{ll} (\frac{R}{\Ra})^{-p} & \;\;\textrm{if}\;\; R \leq
\Ra, \\  (\frac{R}{\Ra})^{-n}  \sqrt{1-\frac{3}{4}(\frac{R-\Ra}{\Ro-\Ra})^2} & \;\;\textrm{if}\;\;
\Ra < R \leq \Ro 
\end{array} \right.
\end{eqnarray}
with the surface density power-law index $p\equiv n+(q-3)/2$ and
the mean surface density at $\Ra$ defined as
\begin{eqnarray}
\Sigma_t &\equiv& 100 \sqrt{2 \pi} \rhoa \ha \Big[\textrm{Erf}(1)+ A\, K(f) \Big] \nonumber \\
&\approx& 21.6 \, \Big(\frac{\rhoa}{3.6\times10^{-16}\,\textrm{g cm}^{-3}}\Big) \Big(\frac{\ha}{16 \,\textrm{au}} \Big)
\Big[\textrm{Erf}(1)+ A \, K(f) \Big]
\;\textrm{g cm}^{-2}
\end{eqnarray} 
where $\textrm{Erf}$ is the error function with $\Erf(1) \approx 0.843$ and
\begin{equation}
K(f) \equiv \frac{\Erf(\sqrt{1+1/f^2})}{\sqrt{1+1/f^2}} \approx f-\frac{f^3}{2} \,.
\end{equation}
Then to check for GI, the Toomre Q parameter of the disk can be derived with \citep{Toomre1964}
\begin{equation}
Q (R) \approx \frac{c_s \Omega}{\pi G \Sigma(R)}
\end{equation}
where $\Omega$ is the angular rotation speed of the disk 
\begin{equation}
\Omega =\sqrt{ \frac{G M(R)}{R^3}}
\end{equation}
with $M(R)$ being the total mass (including that of the central protostar and the disk)
within $R$ 
and $c_s$ is the isothermal sound speed
\begin{equation}
c_s =\sqrt{\frac{k \bar{T}(R)}{\mu m_H}} 
\label{eq:cs}
\end{equation}
where $\mu=2.33$ for molecular gas with H$_2$ and Helium, and $\bar{T}(R)$ is 
the mean temperature at given $R$.
Since the temperature decreases from the midplane to the surface,
the mean temperature at given $R$ can be derived by averaging the density
weighted temperature over $z$ as follow
\begin{eqnarray} 
\bar{T}(R) &=&  \frac{\int \int \rho^s(R,z) T^s(R,z) \frac{d\phi}{2\pi}dz}{\int \int \rho^s(R,\phi, z) \frac{d\phi}{2\pi} dz} \nonumber \\
&\approx &
\Ta (\frac{R}{\Ra})^{-q}  \frac{\Erf(1) + (A+B) K(f) + A\,B K(f/\sqrt{2})}
{\Erf(1) + A K(f)}
\end{eqnarray}
Thus, we have 
\begin{equation}
Q(R) \approx \sqrt{\frac{\bar{T}(R)}{33\, \textrm{K}}\, \frac{M(R)}{1.8 \solarmasse} 
\,\Big(\frac{160 \;\textrm{au}}{R}\Big)^3} 
\,\Big(\frac{21.6\, \textrm{g cm}^{-2}}{\Sigma(R)}\Big)
\end{equation}
Hence, we can constrain the density parameter $\rhoa$ assuming $Q \sim 1$ at $\Ra$.

The mass (including gas and dust) of the disk can be derived by integrating the mean surface density over $R$:
\begin{equation}
M_d  \sim \int_0^{\Ro} \Sigma(R)\, 2 \,\pi \,R dR =  2 \pi C \,\Ra^2 \Sigma_t 
\approx 0.39 \,C \Big(\frac{\Ra}{160 \,\textrm{au}}\Big)^2 \Big(\frac{\Sigma_t}{21.6\, \textrm{g\,cm}^{-2}}\Big) \; \solarmasse
\end{equation}
with
\begin{equation}
C \equiv \frac{1}{2-p}
+\int_1^{\xo} x^{-n+1} \sqrt{1-\frac{3}{4}(\frac{x-1}{\xo-1})^2} dx
\end{equation}
where $\xo \equiv \frac{\Ro}{\Ra}$. Estimation of the disk mass allows us to
check if the disk is massive enough to be gravitationally unstable. 
Moreover, for a disk in vertical hydrostatic equilibrium, the scale height is expected to be
\begin{equation}
\hd^e \sim \frac{c_s}{\Omega} \approx 17.3
\sqrt{\frac{\bar{T}(R)}{33 \textrm{K}}\, \frac{1.8 \solarmasse}{M(R)} \,\Big(\frac{R}{160 \;\textrm{au}}\Big)^3}
\;\textrm{au}
\end{equation}
This expected scale height can also be checked against the best-fit value for consistency.




%

\subsection{Fiducial Disk Model}

The fiducial model with parameters appropriate for the HH 111 disk can be
constructed by fitting the observed total intensity of the dust emission. 
For simplicity, we assume that the dust emission is purely thermal and the
disk is everywhere in LTE.  We use the radiative transfer code reported
in \cite{Lee2021Pdisk} to calculate the dust emission from the model and fit
(by eye) it roughly to the observed maps.

\begin{figure} [!hbp]
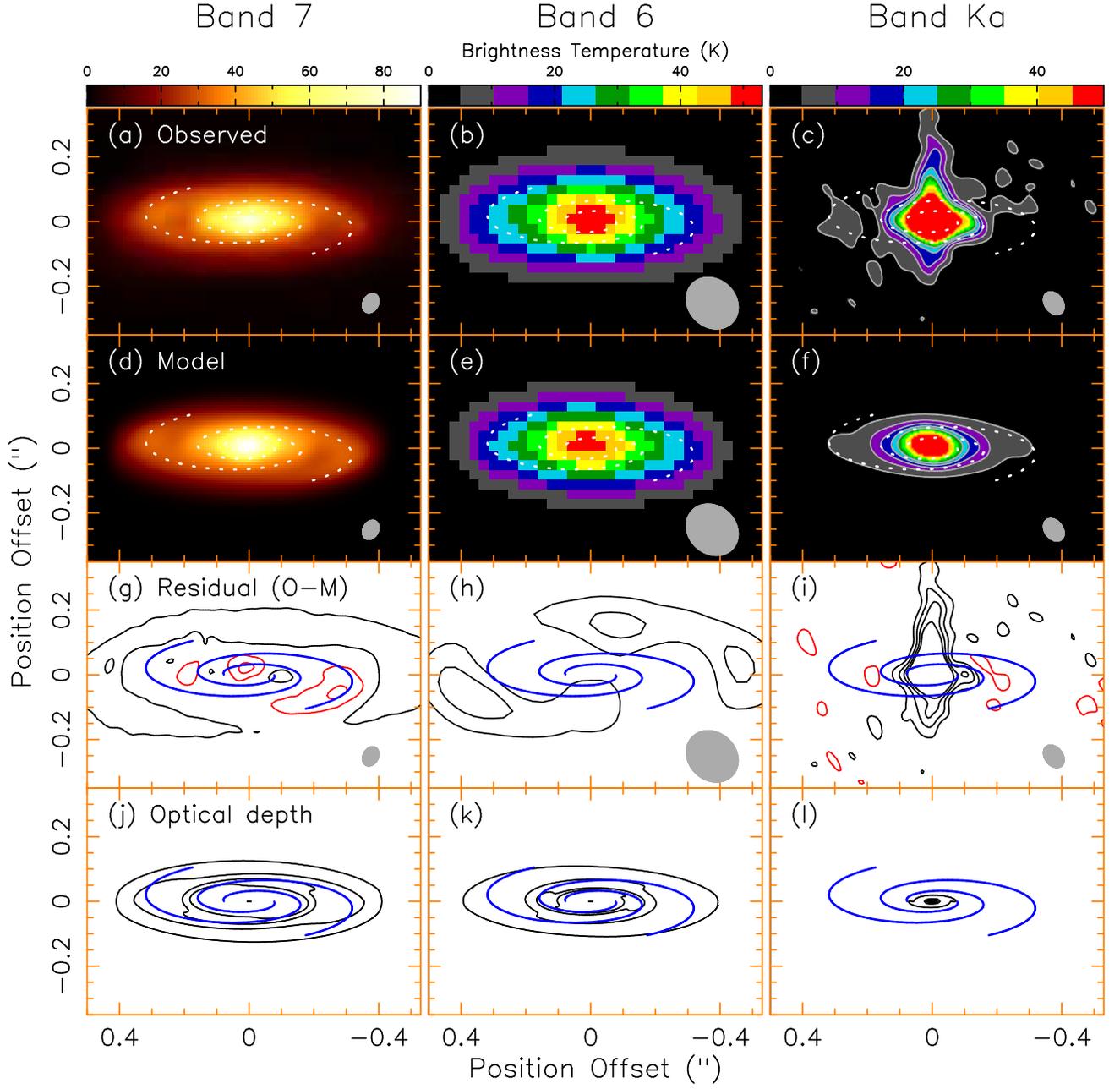

\centering
\putfiga{0.9}{270}{f3.eps} 
\figcaption[]
{Observed maps (1st row), model maps (2nd row), residual maps (3rd row, observed maps $-$ model maps),
and optical depth maps (4th row) in Bands 7 (1st column), 6 (2nd column), and Ka (3rd column).
In (g)-(i), black contours are for positive residuals and red contours for negative residuals.
In (g), contours start at 10 $\sigma$ with a step of 7 $\sigma$, where $\sigma=290$ mK.
In (h), contours start at 20 $\sigma$ with a step of 10 $\sigma$, where $\sigma=110$ mK.
In (i), contours start at 2.5 $\sigma$ with a step of 2.5 $\sigma$, where $\sigma=2.0$ K.
Optical depth starts from 1 with a step of 2.
\label{fig:obsMB}}
\end{figure}

A major uncertainty to calculate the dust emission in the model is the dust
absorption opacity.  Fortunately, since the disk is mostly optically thick
as found later, we can constrain the temperature in the disk with the
observed brightness temperature.  Then, since the spiral arms appears to be
excited by GI \citep{Lee2020HH111}, the Toomre Q parameters there are
required to be $\sim$ 1-2 \citep{Durisen2007}, allowing us to constrain the
density of the disk and thus the dust absorption opacity to achieve the
required optical depth \citep{Beckwith1990,Lin2021}.  Moreover, this disk
has also been mapped in ALMA Band 6 at $\lambda \sim 1.3$ mm (or 231 GHz) at
$\sim$ \arcsa{0}{16} resolution (ALMA archive Project Code: 2016.1.00389.S)
and VLA Band Ka at $\lambda \sim 9.1$ mm (or 32.9 GHz) at $\sim$
\arcsa{0}{07} resolution \citep{Tobin2020} (see Figure \ref{fig:obsMB},
first row).  In Band 6, the disk emission was observed at a longer
wavelength and thus should be less affected by dust self-scattering than
that in Band 7.  In Band Ka, although free-free emission is seen forming a
collimated jet extending out from the inner disk, the emission in the outer
disk should arise mainly from the thermal dust emission.  As can be seen,
the emission there also arises from around the spiral arms.  With multi-band
observations, we can also define an opacity law $\kappa_\lambda = \kappa_0\,
(9.1\,\textrm{mm}/\lambda)^\beta$, where $\kappa_0$ is the dust opacity in
Band Ka at $\lambda \sim$ 9.1 mm and $\beta$ is the dust opacity spectral
index.  Then by fitting the multi-band observations simultaneously, we not
only can better constrain the model parameters in the disk, but also can
determine the parameters in the dust opacity law to constrain the grain
properties.

\begin{figure} [!hbp]
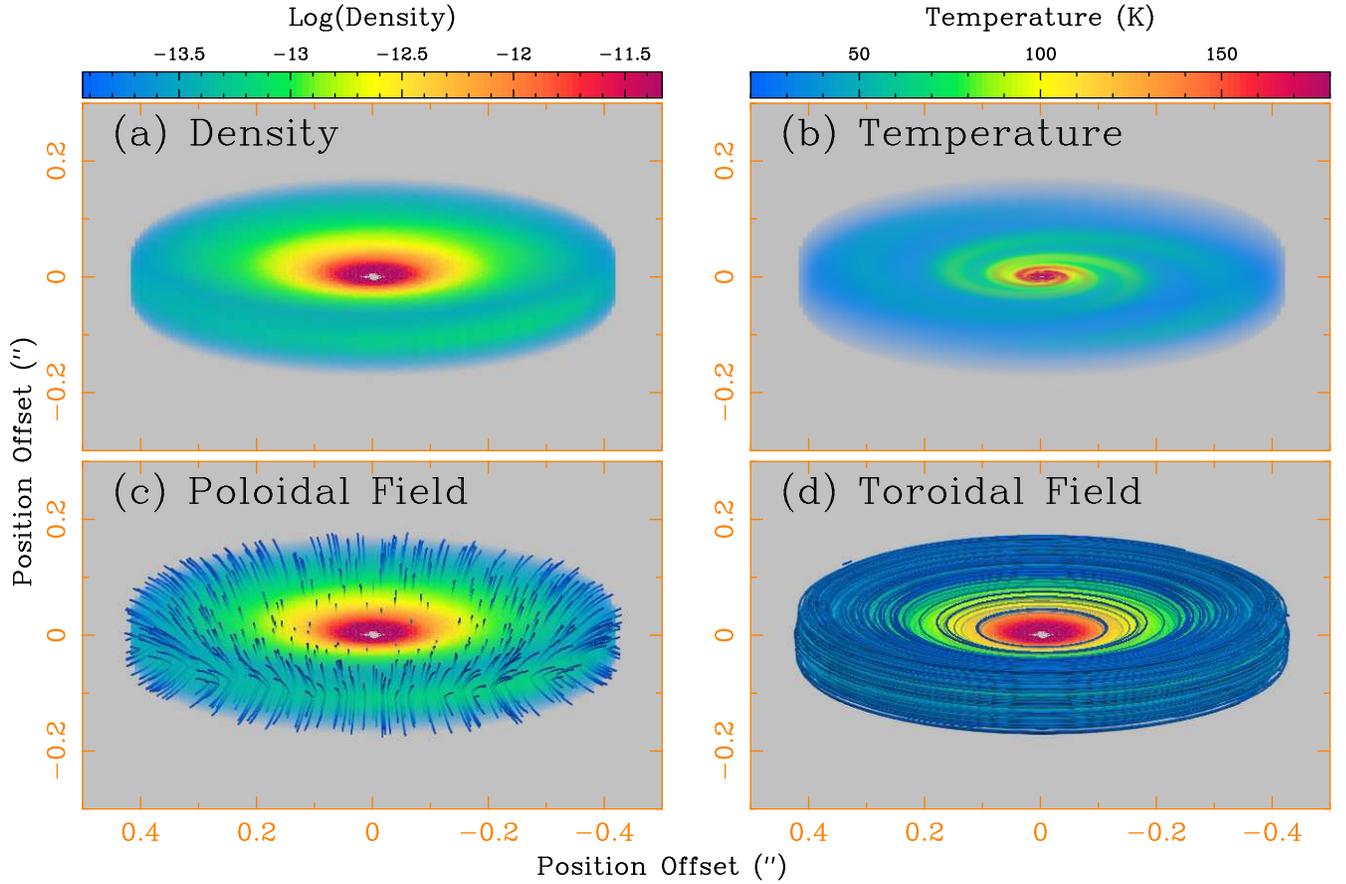

\centering
\putfiga{0.7}{270}{f4.eps} 
\figcaption[]
{Fiducial model for the disk tilted by 18\degree{} away from being edge-on,
as seen for the HH 111 disk.  (a) and (b) show the volume rendering of the
density and temperature distributions in the model, respectively.  See text
for the explanation.  (c) Highly pinched poloidal fields added to the disk. 
(d) Toroidal fields added to the disk.
\label{fig:cartoon}}
\end{figure}

For a geometrically thin accretion disk like the HH 111 disk, we have $q\sim
0.5-0.75$ and $p\sim 1-1.5$ \citep{Armitage2015}.  Thus, we have $n\sim
2.125-2.75$.  In order to simplify our fitting, we assume $n=2.5$.  However,
we allow the value of $q$ to be obtained by fitting the distribution of the
observed brightness temperature in the disk, which is mostly optically thick
in Bands 7 and 6.  Then, as shown in Figure \ref{fig:obsMB}, we find that a
model with $\Ra \sim $ \arcsa{0}{40}, $\ha \sim $ \arcsa{0}{04}, $\Ro \sim $
\arcsa{0}{42}, $\Ta \sim 26$ K, $q \sim 0.65$, $\rhoa \sim
3.6\times10^{-16}$ g \cmc{}, $a \sim -0.22$, $R_s \sim$ \arcsa{0}{40}, $A
\sim 0.5$, $B \sim 0.5$, $f \sim 0.4$, $\kappa_0 \sim 0.12$ \cms{} g$^{-1}$,
and $\beta \sim 1.4$ can roughly reproduce the observed disk emission in
Bands 7 and 6, and the spiral structures in Bands 7 and Ka, except for the
inner disk and jet in Band Ka.  Figures \ref{fig:cartoon}a and
\ref{fig:cartoon}b show the resulting density and temperature distributions
of the disk, produced by the visualization application ParaView
using volume rendering \citep{Ahrens2005}.  Note that with the selected temperature range, we can
see through the disk to the spiral arms in the disk midplane.  However, with
the selected density range, we can only see the overall density structure of
the disk but can not clearly see the spiral arms because 
the density contrast between the spiral arms and
interarms is much smaller than the density range in the disk.
In Bands 7 and 6, the
faint residual emission seen around the disk arises from the inner envelope
around the disk and thus can not be reproduced from our model that only
includes the disk.  Note that the spiral parameters $a$ and $R_s$ are in
good agreement with those previously reported in \citet{Lee2020HH111}.  The
value of $\rhoa$ is set to have Toomre Q parameter of $\sim$ 1 at $\Ra$, so
that the resulting Toomre Q parameter is $\sim$ 1-2 for most of the disk
from $R \sim$ \arcsa{0}{1} to $\Ro$, as required for the disk being
gravitational unstable.  The resulting $\kappa_\lambda$ values are 3.2, 1.8,
0.12 cm$^2$ per gram of dust at $\lambda=$ 0.87, 1.3, and 9.1 mm,
respectively.  In addition, the value of $\kappa_\lambda$ at $\lambda=$ 0.87
mm is similar to those adopted in the Class 0 disks, e.g., in HH 211
\citep{Lee2023HH211} and HH 212 \citep{Lin2021,Lee2021Pdisk}, the value of
$\kappa_\lambda$ at $\lambda=$ 1.3 mm is similar to those adopted in the
Class 0/I disks, e.g., L1527 \citep{vHoff2023} and R CrA IRs7B-a
\citep{Takakuwa2024}, and the value of $\kappa_\lambda$ at $\lambda=$ 9.1 mm
is similar to those adopted in the Class 0/I disks \citep{Tobin2020}.  The
resulting mean dust surface density $\Sigma_t/100$ is $\sim 0.22$ g
cm$^{-2}$ at $\Ra$, and thus the dust emission of the disk is mostly
optically thick in Bands 7 and 6, and mostly optically thin in Band Ka.  The
resulting disk mass $M_d$ is $\sim$ 0.58 \solarmass{}.  Since the protostar
and disk were previously found to have a total mass of $\sim$ 1.8
\solarmass{} \citep{Lee2016HH111}, the protostar has a mass of $\sim$ 1.22
\solarmass{}.  Therefore the disk to protostar mass ratio is $\sim$ 0.48,
and thus the disk is massive enough to excite GI for 2 prominent spiral arms
\cite[see, e.g.,][]{Dong2015}.  The expected scale height in vertical
hydrostatic equilibrium is $\ha^e \sim$ 17.3 au, similar to the best-fit
value of $\ha$, suggesting that the disk is in vertical hydrostatic
equilibrium.  The scale height to radius ratio at $\Ra$ is $\sim$ 0.1,
indicating that the disk is geometrically thin and much thinner than the
younger disks in, e.g., HH 212 \citep{Lee2021Pdisk} and HH 211
\citep{Lee2023HH211}.  In summary, the obtained model parameters are
reasonable and thus the model can be considered as a fiducial model for our
study of dust polarization.

\subsection{Magnetically Aligned Grains}

Given the fiducial disk model, we explore if the dust polarization in the
disk in Band 7 can be due to magnetically aligned grains.  Since this
polarization mechanism only affects slightly the emission intensity by about
the polarization fraction, which is $\sim$ 2\%, no adjustment of the model
parameters is needed.  We use the radiative transfer code reported in
\cite{Lee2021Pdisk}, in which dust polarization by magnetically aligned
grains is included with a parameter $\alpha$ defining the maximum fraction
of polarization in an optically thin region.  Here $\alpha = 0.1$ is used to
produce the small polarization degree of $\sim$ 2\% as observed in the disk,
which is optically thick in Band 7 and thus expected to have a polarization
fraction significantly reduced from the maximum value.  
Interestingly, similar
$\alpha$ value has been adopted to reproduce the polarization images of HL Tau
observed from ALMA Band 7 \citep{Stephens2023} to VLA Q-Band \citep{Lin2024}.

It it believed that magnetically aligned grains have their long axis
perpendicular to the field lines \citep{Andersson2015}.  Therefore, the dust
polarization orientation is perpendicular to the magnetic field in the
optically thin region.  In the optically thick region, the polarization
fraction drops to zero for an isothermal line of sight.  However, if there
exists a temperature gradient, then the polarization orientation can be
perpendicular to the magnetic field if the temperature increases to the
observer or parallel to the magnetic field if the temperature decreases to
the observer \citep{Yang2017,Lin2020}.  The polarized intensity drops to
zero during the transition from the optically thick to more optically thin
region, producing a polarization gap near $\tau \sim$ 4 \cite[see,
e.g.,][]{Lee2021Pdisk}.
At face value in the optically thin limit (see Figure \ref{fig:obsMB}j),
the observed azimuthal
polarization pattern on the far side (Figure \ref{fig:obsmap}d) could be due
to poloidal fields highly pinched toward the equatorial plane (see Figure
\ref{fig:cartoon}c), while the observed polarization orientation
parallel to the minor axis on the near side could be due to toroidal
fields (see Figure \ref{fig:cartoon}d).

In the following, we present the polarization results of the model for the
two field morphologies separately in order to check these possibilities. 
Figures \ref{fig:obsmods}c and \ref{fig:obsmods}d show respectively the
resulting intensity map and polarization map when the disk is threaded with
highly pinched poloidal fields.  Notice that dashed curves are added to
guide the readers for the spiral arms.  In this case, the region with $\tau
\sim$ 4 (marked by the magenta contour in Figure \ref{fig:obsmods}d) happens
to be along the outer part of the spiral arms, producing clear polarization
gaps there.  The gap along the SW arm on the far side is observed, but
observations at higher resolution are needed to check if the gap along the
NE arm on the near side is also observed.  Outside the polarization gaps at
$\tau$ $\sim$ 4, the polarization orientation is roughly azimuthal as
expected and the polarization gaps at the two disk edges (marked by two
tilted cyan ellipses) are due to the depolarization of mutually orthogonal
polarization along the sight lines towards these regions.  This azimuthal
orientation is also observed on the far side of the disk but not on the
nearside of the disk, where the observed orientation is parallel to the
minor axis.  Interior to the polarization gaps at $\tau$ $\sim$ 4, the
polarization orientations are roughly parallel to the minor axis (which is
expected because of polarization reversal due to dichroic absorption along
optically thick sightlines where the temperature decreases to the observer)
and pointing slightly toward the minor axis when moving away from the major
axis.  Although the observed polarization orientations are also roughly
parallel to the minor axis, they point slightly outward from the minor axis. 
In the model, close to the central source, polarized emission is mainly
produced on the near side.  Similar behavior is also seen in the observation
although with much higher intensity.  More importantly, excluding the
polarization gaps along the spiral arms produced by $\tau$ $\sim$ 4 and the
polarization gaps due to the depolarization of mutually orthogonal
polarization, this model produces polarization gaps (regions marked by the
white ellipses) in between spiral arms and high polarization along the
spiral arms, and is thus opposite to the observation.

Figures \ref{fig:obsmods}e and \ref{fig:obsmods}f show respectively the
resulting intensity map and polarization map when the disk is threaded with
toroidal fields.  As in the case of poloidal fields, polarization gaps are
clearly seen along the spiral arms where $\tau \sim$ 4.  Outside these gaps,
the polarization orientations are mainly parallel to the minor axis and
pointing slightly away from the minor axis when moving away from the major
axis, and thus more like pointing radially outward as expected.  This is
different from the observed polarization orientations, which are almost
exactly parallel to the minor axis on the near side and azimuthal on the far
side.  Interior to the polarization gaps at $\tau \sim$ 4, the polarization
orientations are perpendicular to the minor axis, also different from the
observed, which are almost parallel to the minor axis.

In summary, it is unclear if the observed dust polarization of the disk can
be due to magnetic alignment of the dust grains.  The model with poloidal
fields can account only for the azimuthal polarization orientations on the
far side, while the model with toroidal fields fails to account for most
polarization orientations including those polarization orientations parallel
to the minor axis on the near side.  Adding the toroidal fields into the
model with poloidal fields will tilt the polarization orientations to one
side (not shown), also inconsistent with the observed polarization
orientations.


\subsection{Dust Self-Scattering}

\def\amax{$a_\textrm{\scriptsize max}$}


At our observed wavelength of $\sim$ 870 \micron{}, dust scattering is
efficient when the maximum grain size \amax{} $\sim$ 70-210 \micron{} (with
a peak at $\sim$ 140 \micron) and becomes inefficient when \amax{} increases
to $\gtrsim$ 300 \micron{} \citep{Kataoka2015}.  Recently DSHARP dust
opacities have been used to model the dust emission in protoplanetary disks
\citep{Birnstiel2018}, assuming compact grains with a MRN size distribution. 
For this type of grains with \amax{} $\sim$ 70-210 \micron{},
$\kappa_\lambda$ is $\sim$ 0.8-1.9 cm$^2$ per gram of dust, much smaller
than that required in our fiducial model.  We could reduce $\kappa_\lambda$
in the fiducial model by increasing the density and thus the disk mass. 
However, since the disk mass to protostellar mass ratio is already large in
our model, it is not clear if the disk mass can be increased significantly,
unless the total mass of protostar and disk is found to be higher.  It is
possible that the dust grains in this disk have different composition from
that assumed in DSHARP for older disks where the planet formation has begun. 
Here for demonstration purpose, we adopt the DSHARP grains with \amax{} of
170 \micron{}, which has a not-too-small absorption opacity and is still
efficient in dust scattering.  The dust absorption opacity is $\sim$ 1.4
cm$^2$ per gram of dust but is scaled up to 3.2 cm$^2$ per gram of dust. 
The scattering opacity is $\sim$ 10 cm$^2$ per gram of dust, giving rise an
albedo of $\sim$ 0.76.  Note that this combination could be obtained if
using porous grains \citep{Tazaki2019}.



\begin{figure} [!hbp]
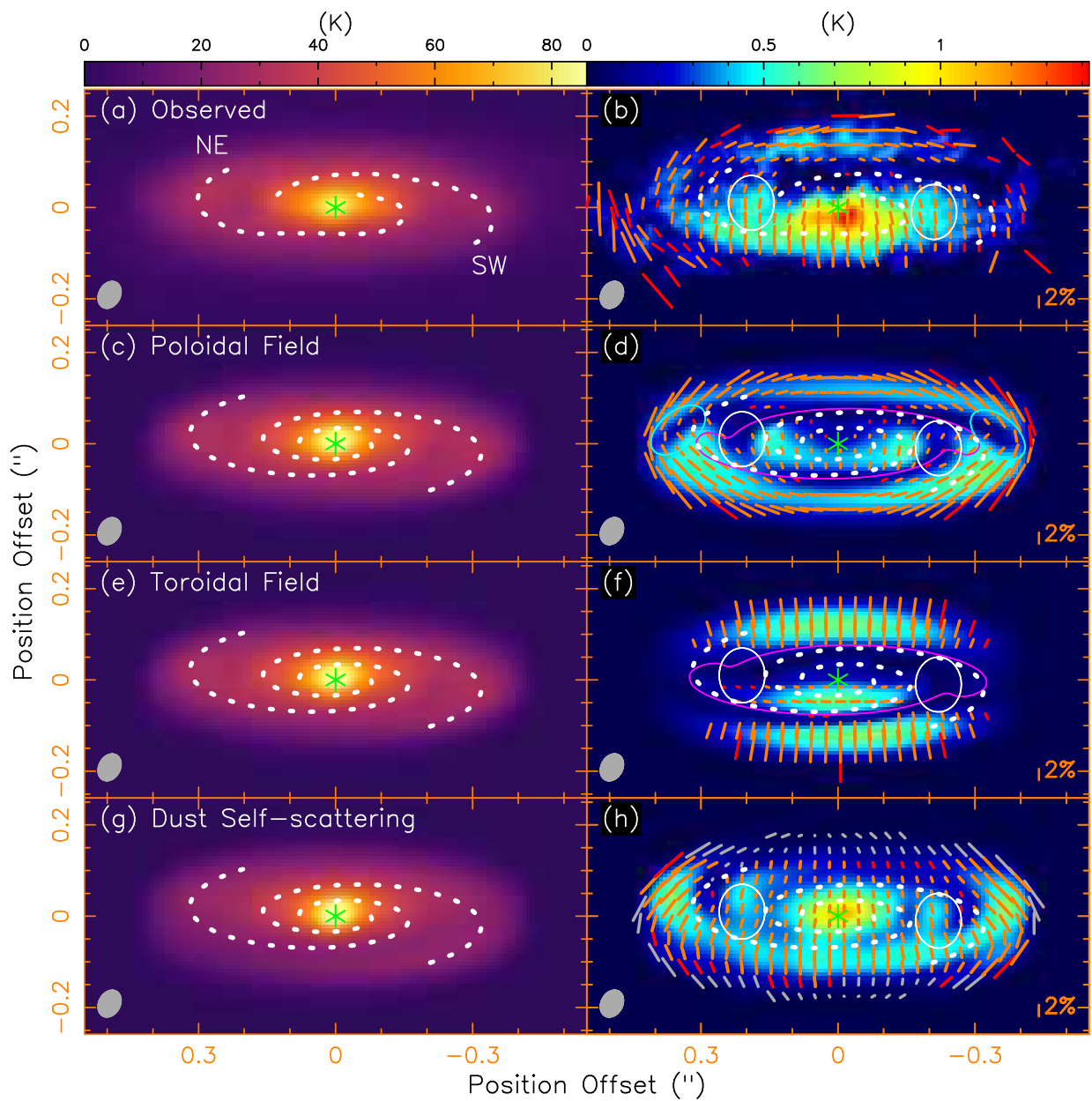

\centering
\putfiga{0.9}{270}{f5.eps} 
\figcaption[]
{Comparison of the polarization between the observed maps (1st row)
and the model maps derived from the magnetically aligned grains by poloidal fields (2nd row) and
toroidal fields (3rd row) and the model map due to
dust self-scattering (4th row). In right column, the white ellipses mark the interarm regions.
In (d), the cyan ellipses mark the regions where the polarization gaps are due to depolarization
of mutually orthogonal polarization. In (h), gray line segments are added to show the detections
below $3\sigma$.
\label{fig:obsmods}}
\end{figure}

We use RADMC-3D \citep{Dullemond2012} to calculate the dust polarization due
to dust self-scattering.  Since the dust scattering causes an additional
extinction to the disk emission, the disk temperature is increased by a
factor of $\sim$ 1.3 in order to roughly match the observed brightness
temperature in the disk, as shown in Figure \ref{fig:obsmods}g.  Figure
\ref{fig:obsmods}h shows the polarized intensity, orientations, and degree
generated from the model.  The polarization degree is $\sim$ 2\%, roughly
consistent with the observation.  The polarization orientation is mainly
parallel to the minor axis pointing slightly away from the minor axis and
becomes azimuthal at the two disk edges in the major axis, as seen in the
self-scattering model presented in \citet{Yang2017} when the disk is
optically thick.  The model produces bright polarized emission at the lower
disk rim and the two disk edges, as seen in the observation, due to a sharp
decrease of the disk density there in our model, creating an asymmetry in
the flux toward the lower rim and two edges.  However, the model produces
very faint polarized emission at the upper rim, inconsistent with the
observation in which bright polarized emission is detected with azimuthal
orientation.  Adding the line segments (gray line segments) for
polarized intensity at lower level, we can see that polarization
orientations in the upper rim (where the emission is optically thin) are
also quite different from being azimuthal.

More importantly, we find that the polarized intensity is lower along the
spiral arms, but higher in the interarms (as marked by the two white
ellipses) between the spiral arms.  This behavior is better seen along the
major axis, because the incident light along the major axis is scattered by
90\degree{} into the line of sight and is thus maximally polarized
\citep{Yang2017}.  This distribution is also seen in the observed
polarization map.  The detailed distribution and intensity of the polarized
emission are different from the observed polarization map because the actual
spirals may have different physical properties (e.g., temperature and
density) from those used in our simple model and the actual dust grains are
different from the DSHARP grains.  The interarms are sandwiched and
illuminated by two brighter spiral arms and thus have higher polarized
intensity.  Going inward to around the central source, no clear near-far
side asymmetry is seen in the polarized intensity, inconsistent with the
observed.  Such a near-far side asymmetry can be produced with a
geometrically thick disk viewed at a moderate inclination angle
\citep{Yang2017}.  However, since the disk in our model is highly inclined,
no clear near-far asymmetry can be seen, even if we increase the scale
height $h_t$ by 30\% (the uncertainty in our model) to have a geometrically
thick disk.





In summary, the dust self-scattering model can roughly reproduce most of the
observed polarization features including the polarization orientations
parallel to the minor axis on the near side, the azimuthal polarization
orientations near the disk edges, and the anticorrelation of the polarized
intensity with the spiral arms (i.e., lower polarized intensity in between
spiral arms and higher the lower polarized intensity along the spiral arms). 
Therefore the observed dust polarization in the HH 111 disk is likely
dominated by dust self-scattering, with possible contributions from other
mechanisms, such as grains aligned by a poloidal magnetic field.
The polarization orientations parallel to the minor
axis and the azimuthal polarization orientations near the disk edges have
also been seen in other disks and attributed to the dust self-scattering
\cite[e.g.,][]{Bacciotti2018,Girart2018}.  Thus, our result indicates a
grain growth in the disk to a size of $\sim$ 150 \micron{}.  The best-fit
dust opacity spectral index $\beta \sim 1.4$ also supports this grain growth
possibility.

The anticorrelation of the polarized intensity with the spiral arms is
reminiscent of that seen in HL Tau \citep{Stephens2023}, where rings are
observed to be less polarized than the gaps.  The interpretation for the HL
Tau case was that the polarizations in the rings and gaps are both produced
by azimuthally aligned effectively prolate grains through both thermal
emission and scattering.  The lower optical depth in the gaps allows the
thermal emission of the aligned grains (which has a much higher intrinsic
polarization fraction) to dominate over scattering, producing a relatively
high polarization fraction.  The higher optical depth in the rings reduces
the thermal polarization through dichroic extinction to such an extent that
it becomes dominated by scattering, which tends to produce a relatively low
polarization fraction.  It is plausible that the polarization observed in
our case of HH 111 is also produced by scattering aligned grains.  In
particular, an azimuthally aligned effectively prolate grains would readily
produce the azimuthal pattern observed in the upper (far) side of the outer
disk outside the SW spiral through thermal emission, if the optical depth
there is relatively low such that thermal emission by aligned grains can
dominate scattering as well as part of the polarization parallel to the
minor axis observed in the lower (near) side of the inner disk from
polarization reversal due to dichroic extinction if the optical depth there
is large enough (see Fig.  5d, also Fig.  9 in Lin et al.  2020).  In the
lower (near) side of the disk, scattering could be enhanced compared to that
in the far side if the disk is geometrically thick enough for the $\tau=1$
surface on the near side to become significantly more inclined to the line
of sight than that on the far side \citep{Yang2017}.  With potential boost
from this geometric effect, scattering could dominate over polarized thermal
emission over most of the disk (except near the edge on the far side) and
broadly explain the main observed features, particularly the weaker
polarization observed in the spiral arms, as we have demonstrated through
detailed modeling (see Fig.~5h).  In this scenario, the main difference
between HH 111 and HL Tau is that the former is more inclined and has a
geometrically thicker dust layer responsible for the polarized submillimeter
emission, which increases the scattering-induced polarization on the near
side relative to that on the far side, and a larger optical depth along the
line of sight (boosted by a more edge-on disk inclination), which further
enhances the minor-axis aligned polarization on the near side of the disk by
flipping the orientation of the polarization emitted by the azimuthally
aligned effectively prolate grains by 90\degree{} through dichroic
extinction. 
The flipping is more likely on the near side of the disk than the far
side because, for highly inclined disks such as HH111, the polarized thermal
emission from aligned grains near the midplane passes through a larger
column of dust (and gas) along the line of sight on the near side than the
far side (as illustrated by the cartoon in Figure 11 of Takakuwa et al. 
2024); the flipped polarization adds to (rather than cancel) the
polarization from scattering at locations along the minor axis.
Another potential difference is that the scattering
(nonspherical) grains in the HH 111 disk could be aligned by a (pinched)
poloidal field, which can broadly explain the azimuthal polarization pattern
(of non-scattering origin) observed near the disk edge on the far side (see
Fig.~5d).  
Detailed calculations that include self-consistent treatment of
both thermal emission and scattering by aligned grains,
particularly optical depth effects such as dichroic extinction (which
may be different on the near and far side of the disk),
 are needed to test
the scenario quantitatively.  Similar to the work on HL Tau, additional
observations at longer wavelengths can further distinguish the origin of
polarization.  Since polarization from scattering decreases at longer
wavelengths and that from aligned grains increases, longer wavelength
observations can better reveal the underlying grain alignment configuration.

{\bf

}

\acknowledgements

We thank the anonymous referee for useful comments.
This paper makes use of the following ALMA data: ADS/JAO.ALMA\#
2015.1.00037.S and 2017.1.00044.S.  ALMA is a partnership of ESO
(representing its member states), NSF (USA) and NINS (Japan), together with
NRC (Canada), NSTC and ASIAA (Taiwan), and KASI (Republic of Korea), in
cooperation with the Republic of Chile.  The Joint ALMA Observatory is
operated by ESO, AUI/NRAO and NAOJ.  C.-F.L.  and Y.-C.H.  acknowledge
grants from the National Science and Technology Council of Taiwan
(110-2112-M-001-021-MY3 and 112-2112-M-001-039-MY3) and the Academia
Sinica (Investigator Award AS-IA-108-M01).  ZYL is supported in part by NASA
80NSSC20K0533 and NSF AST-2307199.





\def\nat{Natur}

\appendix



\end{document}